\documentclass[12pt,preprint]{aastex}
\begin{document}

\title{Quasi-Periodic Variability in NGC 5408 X-1} \author{Tod
E. Strohmayer$^1$, Richard F. Mushotzky$^1$, Lisa Winter$^2$, Roberto
Soria$^3$, Phil Uttley$^4$, \& Mark Cropper$^5$}
\altaffiltext{1}{Astrophysics Science Division, NASA's Goddard Space
Flight Center, Greenbelt, MD 20771 email: stroh,
richard@milkyway.gsfc.nasa.gov} \altaffiltext{2}{Astronomy Department,
University of Maryland, College Park, MD 20742;
lwinter@astro.umd.edu}\altaffiltext{3}{Harvard-Smithsonian Center for
Astrophysics, 60 Garden St, Cambridge, MA 02138;
rsoria@head.cfa.harvard.edu}\altaffiltext{4}{Astronomical Institute
`Anton Pannekoek', University of Amsterdam, Kruislaan 403, 1098 SJ,
Amsterdam, the Netherlands}\altaffiltext{5}{Mullard Space Science
Laboratory, University College London, Holmbury St Mary, Dorking,
Surrey RH5 6NT;msc@mssl.ucl.ac.uk}

\begin{abstract}

We report the discovery with XMM-Newton of quasiperiodic variability
in the 0.2 - 10 keV X-ray flux from the ultraluminous X-ray source NGC
5408 X-1. The average power spectrum of all EPIC-pn data reveals a
strong 20 mHz QPO with an average amplitude (rms) of 9\%, and a
coherence, $Q \equiv \nu_0 / \sigma \approx 6$. In a 33 ksec time
interval when the 20 mHz QPO is strongest we also find evidence for a
2nd QPO peak at 15 mHz, the first indication for a close pair of QPOs
in a ULX source. Interestingly, the frequency ratio of this QPO pair
is inconsistent with 3:2 at the 3$\sigma$ level, but is consistent
with a 4:3 ratio.  A powerlaw noise component with slope near 1.5 is
also present below 0.1 Hz with evidence for a break to a flatter slope
at about 3 mHz. The source shows substantial broadband variability,
with a total amplitude (rms) of about 30\% in the 0.1 - 100 mHz
frequency band, and there is strong energy dependence to the
variability. The power spectrum of hard X-ray photons ($> 2$ keV)
shows a ``classic'' flat-topped continuum breaking to a power law with
index 1.5 - 2. Both the break and 20 mHz QPO are detected in the hard
band, and the 20 mHz QPO is essentially at the break. The QPO is both
strong and narrow in this band, having an amplitude (rms) of $15 \%$,
and $Q \approx 25$.  The energy spectrum is well fit by three
components, a ``cool'' disk with $kT = 0.15$ keV, a steep power law
with index 2.56, and a thermal plasma at $kT = 0.87$ keV. The disk,
power law, and thermal plasma components contribute 35, 60, and 5 \%
of the 0.3 - 10 keV flux, respectively.  Both the timing and spectral
properties of NGC 5408 X-1 are strikingly reminiscent of Galactic
black hole systems at high inferred accretion rates, but with its
characteristic frequencies (QPO and break frequencies) scaled down by
a factor of 10 - 100.  We discuss the implications of these findings
in the context of models for ULXs, and their implications for the
object's mass.

\end{abstract}

\keywords{black hole physics - galaxies: individual: NGC 5408 - stars:
oscillations - X-rays: stars - X-rays: galaxies}

\section{Introduction}

The bright X-ray sources found in nearby galaxies, the ultraluminous
X-ray sources (ULX), have been the focus of considerable study over
the last few years.  The ULXs observed to date span a range of
luminosities from several $\times 10^{39-41}$ ergs s$^{-1}$. To
summarize, their implied isotropic luminosities may be incompatible
with the Eddington luminosity of accreting black holes (BHs) whose
upper mass is set by the evolution of ``normal'' stars (by this we
mean all but the earliest populations of stars).  There are basically
three solutions to this luminosity conundrum; either the objects are
intermediate-mass BHs (Colbert \& Mushotzky 1999), or, they
are stellar BHs with, in some cases, substantial beaming of
their X-ray radiation (King et al. 2001); or, they are stellar-mass
BHs emitting above their Eddington limit (Begelman 2006).  It is
possible that some ULXs appear very luminous because of a combination
of all three factors (moderately higher mass, mild beaming and mild
super-Eddington emission). It may also be that the ULXs are an
inhomogeneous population, comprised of both a sub-sample of
intermediate-mass BHs and moderately beamed stellar black
holes (for recent reviews see Fabbiano \& White 2006; Miller \&
Colbert 2004).

Efforts to unambiguously measure the masses of ULXs have been
frustrated by the difficulty in finding counterparts at other
wavelengths. The inability to directly detect binary motion of the
putative companions has so far precluded the use of the familiar
methods of dynamical astronomy to weigh these BHs (see, however,
Kaaret, Simet \& Lang 2006).  Rather, indirect methods based on broad
band spectral and timing measurements have been employed, but these
are still plagued by systematic uncertainties associated with
theoretical interpretation of the observed properties. For example,
standard accretion disk theory predicts that at a given (fixed) value
of $L/L_{Edd}$ the inner disk temperature will scale with mass as $T
\propto M^{-1/4}$ (see, for example, Frank, King \& Raine 2002), thus
cool thermal components (with $T < 0.5$ keV) in the spectra could be
an indication of super-stellar masses.  Miller and colleagues found
several examples of cool disk components in the 0.1 - 0.2 keV range,
most notably in the two bright ULXs in NGC 1313 (Miller et al. 2003).
Comparisons with Galactic systems and the expected mass scaling give
mass estimates in the $M \sim 10^{2-3}$ range.  Subsequent work has
found similar cool components in additional objects (Winter, Mushotzky
\& Reynolds 2006; Kaaret at al.  2003; Miller, Fabian \& Miller 2004a;
Miller, Fabian \& Miller 2004b; Dewangan et al. 2004; Cropper et
al. 2004; Kong, DiStefano \& Yuan 2004; Stobbart, Roberts \& Wilms
2006; and Feng \& Kaaret 2005).

In addition to spectroscopy, X-ray timing observations of ULXs have
increasingly been sought in an attempt to derive mass constraints
based on comparisons with both the Galactic population of stellar BHs
and the extragalactic AGNs.  There is good evidence from comparison of
Fourier power spectra of AGN and stellar BHs that for similar
accretion states characteristic time-scales (such as power spectral
break frequencies) scale with black hole mass (see, for example,
McHardy et al. 2006; Markowitz et al. 2003). Moreover, the detailed
timing properties of several stellar BHs with good mass measurements
can, in principle, serve as a ``calibration standard'' for estimating
masses of ULXs, if similar timing signatures can be detected in them.
Recent successes in this area include the detection with XMM-Newton of
quasi-periodic oscillations (QPOs) in the 50 - 100 mHz range from a
bright ULX in M82 (Strohmayer \& Mushotzky 2003; Mucciarelli et
al. 2006; Fiorito \& Titarchuk 2004), and subsequent detection of a
break frequency (Dewangan, Titarchuk \& Griffiths 2006; Mucciarelli et
al. 2006). Although it seems likely that both the break and QPO are
associated with the same ULX in M82, it has not been possible to prove
this based on XMM-Newton imaging alone (see Kaaret, Simet \& Lang 2006
for a discussion).

Several other ULXs have been reported to show evidence of X-ray timing
signatures.  Cropper et al. (2004) reported evidence for a break in
the Fourier power spectrum of NGC 4559 X-7 at about 30 mHz based on
XMM-Newton observations, though they also could model the spectrum
with a QPO rather than a break.  This source also shows evidence for a
soft thermal component at 0.12 keV, similar to the ``cool disk''
components in other ULXs.  They argued that both the spectral and
timing results provide support to an intermediate-mass black hole
interpretation.  Soria et al. (2004) studied the ULX NGC 5408 X-1 with
a series of relatively short XMM-Newton pointings.  Like other ULXs,
the source shows a cool thermal component ($kT \approx 0.14$ keV), but
it also has a steep-spectrum radio counterpart (Kaaret et al. 2003).
The radio detection has been used to argue that relativistically
beamed jet emission might explain both its radio and X-ray emission.
More recent radio data suggest that the radio emission is not
analagous to the flat spectrum core emission seen in Galactic black
holes in the low-hard state, but is more likely a radio lobe powered
by jet emission from the hole (Soria et al. 2006). From the short
XMM-Newton exposures it is evident that the source is quite variable,
and Soria et al. (2004) reported evidence for a break in the power
density spectrum at about 3 mHz.  Based on the strong varibility and
the need to confirm the break frequency, we proposed a longer
XMM-Newton pointing at this source.  In this paper we report the
initial findings of these observations.

\section{XMM Observations and Data Analysis}

XMM-Newton observed NGC 5408 for $\approx 130$ ksec beginning on
January 5, 2006 at 19:03:57 (TT).  For our study we used only the EPIC
data. We used the standard SAS version 7.0.0 tools to filter and
extract images and event tables for both the pn and MOS cameras. We
detect the source easily and there are no source confusion
problems. We extracted events in an 18'' radius around the source in
both the pn and MOS cameras. The observation was affected by
relatively high background rates (flaring) only for relatively brief
periods near the beginning and end of the pointing. We were able to
produce lightcurves in three intervals of 34, 56, and 17 ksec lengths,
for a total useful exposure of about 107 ksec. Figure 1 shows a
combined pn+MOS lightcurve (150 s bins) including photons over the
full bandpass (0.2 - 15 keV) of each instrument.  The mean count rate
is about 1 s$^{-1}$, and even by eye one can deduce that the source is
varying significantly.

\subsection{Power Spectral Timing Analysis}

Since the countrate is higher in the pn, and it has a higher sampling
rate, we began our Fourier analysis with the pn data.  We broke each
good interval into segments of length 16.5 ks, constructed time series
sampled at 512 Hz, and calculated the power spectrum for each
interval.  This gave a total of 6 power spectra, each with a lower
frequency bound of $6.06 \times 10^{-5}$ Hz, low enough to sample the
putative break frequency reported by Soria et al. (2004).  The average
power spectrum rebinned to 1.5 mHz resolution is shown in Figure
2. All power spectra shown here use the so called Leahy normalization,
with the poisson noise level being 2 (Leahy et al. 1983). The spectrum
rises below 0.1 Hz indicating the presence of significant variability,
and we find a candidate QPO peak evident near 20 mHz.  To further
quantify the variability we fitted a model to the power spectrum. For
the continuum we used a broken power-law plus a constant.  While this
model is adequate below about 8 mHz and above 50 mHz, it is clearly
inadequate in the range between these limits.  We find a statistically
acceptable fit with the inclusion of three additional Lorentzian
components.  Fitting the power spectrum up to 0.3 Hz we find a best
fit with $\chi^2 = 102.2$ for 118 degrees of freedom (dof, 132 data
bins, and 14 parameters).  This model fit is also shown in Figure 2
(thick solid curve).  A summary of the model parameters is given in
Table 1.

The strongest Lorentzian component is the QPO feature at 20 mHz. Our
fit gives a centroid of $19.8 \pm 0.2$ mHz, a width of $3.4 \pm 1$
mHz, and an amplitude (rms) of $9 \pm 2 \%$.  If we exclude this
component the $\chi^2$ value increases by 35.1. Using the F-test to
estimate the significance of the change in $\chi^2$ associated with
the 3 additional parameters for the 20 mHz QPO we find a chance
probability of $1.3 \times 10^{-7}$, which is a bit better than a
5$\sigma$ detection.  The two other Lorentzian components are not as
strongly required. We estimate F-test probabilities for the 11 mHz and
28 mHz components of $4.5 \times 10^{-2}$, and $9 \times 10^{-4}$,
respectively.  Based on our modeling of this spectrum alone we
conclude there is good evidence for a 20 mHz QPO, with some evidence
for additional variability components in the 8 - 30 mHz range. In
particular the apparent need for the higher frequency component is
related to the rather steep rise in the power spectrum below 40 mHz.
While no sharply peaked QPO bump is evident here, the quick rise in
Fourier power is significant.  Whether this is truly a QPO, or part of
a more complex variability continuum is difficult to say based solely
on this power spectrum.

Our model fit gives a break frequency, $\nu_b = 3.5 \pm 0.4$ mHz, that
appears to be consistent with that initially reported by Soria et
al. (2004) based on shorter observations.  To assess the need for the
break in the power-law component we also modelled the power spectrum
with only a single power-law but including the three Lorentzian
components.  This model does not fit very well, and $\chi^2$ increases
by more than 40, which is very significant based on the F-test.  We
therefore conclude that the break is real.  The broken power-law
continuum has an integrated amplitude (rms) of $15.5 \%$ in the 0.1 -
1000 mHz frequency band.

Mucciarelli et al. (2006) found that the QPO associated with a ULX in
M82 could vary in frequency on timescales of hours. We separately
examined the power spectra of the three individual good time intervals
to see if the 20 mHz QPO could be detected and if any time dependence
was evident. The longest continuous good interval (interval 2, 56
ksec) shows a particulary strong QPO peak at 20 mHz (see Figure 3).
The other good time intervals show indications of excess power at 20
mHz, but the signal to noise ratio is lower than for interval 2. We
did not find any strong evidence for variation of the QPO frequency.
Interestingly, in this data interval the ``bump'' just above 10 mHz
also becomes more prominent. We fitted the same model as discussed above
and found that exclusion of this Lorentzian component in the fit now
increases $\chi^2$ by 15.3.  This gives an F-test probability of a
little less than 1/1000 against the presence of this component in this
power spectrum.  Indeed, a power spectrum using only the first 33 ksec
of interval 2 shows a pair of rather prominent peaks (see Figure 4).
In this spectrum the statistical requirement for {\it both} QPO peaks
is strong; the F-test yields chance probabilities of $2 \times
10^{-7}$, and $1 \times 10^{-10}$ for the 14 and 20 mHz features,
respectively.  While the evidence for two QPOs in the average of all
the data is relatively modest, it is much stronger when the more
robust 20 mHz feature is also more prominent.  Since we did some
additional data selections when looking at the 2nd good time interval,
the above probabilities should be increased by a trials penalty, but
even using a factor of ten (which is conservative), we still have
rather strong statistical indications for a pair of QPOs during this
interval.  We suggest that NGC 5408 X-1 can sometimes show a pair of
sharp, closely spaced QPOs. We will have more to say about the
possible implications of this shortly.

We also explored the energy dependence of the variability by creating
power spectra in two broad bands, 0.2 - 2 keV, and $ > 2$ keV.  In
order to increase the count rates we used both the pn and MOS for this
study.  Figure 5 shows the average power spectrum in the hard band for
all the data.  A narrow feature at 20 mHz is readily apparent, and the
overall power spectral shape is strikingly reminiscent of the
``flat-topped'' noise with a QPO and break commonly seen in many
Galactic systems (see for example, McClintock \& Remillard 2006).  We
find that this power spectrum is well fitted by a broken power-law
continuum, but not a single power-law, and we derive a break frequency
of $25.2 \pm 5$ mHz.  The band limited noise component has an
amplitude (rms) of $24 \%$.  Figure 6 compares the power spectra in
the hard and soft bands.  Interesting energy dependences are evident,
with the soft flux contributing more variability power at low
frequencies, with the result that the derived break frequency and the
slope above the break both depend on energy.  An effect like this has
also been seen in Galactic BHs (Belloni et al. 1997; Reig, Belloni \&
van der Klis 2003).  The QPO peak at 20 mHz is remarkably sharp in
this energy band, we derive a coherence, $Q \equiv \nu_0
/\sigma_{lore} = 25$, and an amplitude (rms) of $15 \%$.  Comparison
of this amplitude with that in the full band shows that the QPO
amplitude increases with energy, a common feature of QPOs in Galactic
binaries as well.

\subsection{Energy Spectral Analysis}

As noted in the Introduction, previous spectral studies have shown
that NGC 5408 X-1 has a cool thermal component with $kT \approx 0.14$
keV (Kaaret et al. 2003; Soria et al. 2004).  We obtained a pn
spectrum by extracting an 18'' region around the source.  Background
was obtained from a nearby circular region free of sources.  We first
fitted a model including emission from a relativistic disk ({\it diskpn}
in XSPEC) and a power-law.  These components were modified by
successive photoelectric absorption components, one fixed at the best
Galactic value ($n_H = 5.7 \times 10^{20}$ cm$^{-2}$), the other left
free to account for possible local absorption.  Interestingly, this
model does not provide an adequate fit, resulting in $\chi^2 = 818$
with 668 dof.  The nature of the residuals suggest the possibility of
thermal plasma emission, so we added an {\it apec} component and
fitted again.  Including the {\it apec} component results in an
acceptable fit with $\chi^2 = 682.3$ for 666 dof (see Table 2 for a
summary of spectral parameters, and Figure 7 for a plot of the energy
spectrum and best fitting model).  Such a component may be fairly
common in ULXs as this is now one of several sources with such
indications in its spectrum, the others being a recently found object
in NGC 7424 (Soria et al. 2006), and Holmberg II X-1 (Dewangan et al.
2004).  The inferred disk temperature of $0.15 \pm 0.01$ keV, and
power-law index of $2.56 \pm 0.04$ are similar to earlier findings.
The unabsorbed flux is $3.15 \times 10^{-12}$ ergs cm$^{-2}$ s$^{-1}$
(0.3 - 10 keV), and implies an X-ray luminosity of $8.7 \times
10^{39}$ ergs s$^{-1}$ (at a distance of 4.8 Mpc, Karachentsev et
al. 2002).  The power-law, black body, and {\it apec} components
contribute, 60.5, 34.8, and 4.6 \% of the energy flux, respectively.

\section{Discussion and Summary}

Our XMM-Newton observations of NGC 5408 X-1 reveal a wealth of new
X-ray variability phenomena.  We have found strong evidence for a QPO
at 20 mHz, as well as a break in the power spectral continuum near the
QPO frequency.  This is the 2nd ULX to show evidence for both a QPO
and break in its power spectrum, the other being a ULX in M82
(Dewangan et al.  2006). In addition, during the time when the 20 mHz
QPO is strongest, we also find evidence for a 2nd QPO at 14.8 mHz (see
Figure 4). This is the first evidence for a pair of closely spaced QPO
peaks in a ULX. Above about 2 keV the form of the power spectrum is
clearly that of ``flat-topped'' noise breaking to a power-law with
index of order 1.5 - 2, with a QPO at or near the break frequency.
This form is common among Galactic black hole systems, and is most
often associated with sources in either the low-hard state and/or the
very high state (the steep power-law state, in the nomenclature of
McClintock \& Remillard 2006).  The break frequency we derive appears
to depend on the energy band, with the inferred break increasing with
energy.  This behavior is also seen in Galactic systems (see Belloni
et al. 1997; Reig et al. 2003).

What do these results imply for the mass of NGC 5408 X-1?  Both its
X-ray spectral and timing properties are suggestive of a black hole at
a relatively high mass accretion rate (with respect to the Eddington
rate). In terms of a state classification we suggest that an analogy
with either the very high state or perhaps the ``intermediate'' state
is more appropriate than that of a classical low-hard state.
Although low-hard state power spectra often show the ``flat-topped''
power continuum, QPOs are rarer than in the very high state, and the
steep (energy) power-law and high luminosity of X-1 are even more
problematic for a low-hard state analogy.  In addition to this X-ray
evidence, the radio counterpart to X-1 has a steep spectrum that is
inconsistent with the flat spectrum, compact jets seen from Galactic
systems in the low-hard state.

We thus suggest that the 20 mHz QPO reported here is analogous to
those identified in several Galactic systems and whose frequency is
strongly correlated with parameters describing the energy spectrum,
for example, the slope of the power-law component.  Systems which show
such behavior include, XTE J1550-564 (hereafter 1550); GRO J1655-40
(hereafter 1655, Sobczak et al. 2000); GRS 1915+105 (hereafter 1915);
XTE J1748-288; 4U 1630-47 (Muno, Morgan \& Remillard 1999; Vignarca et
al. 2003); and XTE J1859+226 (Casella et al. 2006).  Galactic BHs
often show a complex QPO phenomenology, and several QPO types (dubbed
type-A, -B and -C) have been proposed in the literature (Wijnands et
al. 1999; Homan et al. 2001; Remillard et al. 2002; Casella et
al. 2004; 2006). The strongest (in terms of fractional rms),
narrowest, and most ubiquitous of these types are the type-C QPOs.
Based on its observed properties it appears likely that the 20 mHz QPO
in X-1 is similar to the type-C QPOs in the Galactic systems.  It may
be that the 15 mHz QPO is a type-A or -B QPO.  For example, Casella et
al. (2004) show that in XTE J1859+226 quick transitions can occur
between QPO types. Interestingly, they found that transitions from
type-C to -B occurred, and at that time the QPO frequencies were in a
4:3 ratio.  Perhaps something similar is occuring in X-1, but more
data will be required to confirm this.

Vignarca et al. (2003) compile results from several of the objects
noted above, and the strong correlation between energy spectral index
and QPO frequency is evident in their Figure 10. If a similar
correlation holds for X-1 then one can infer an expected QPO frequency
based on the measured spectral index, and deduce a simple mass scaling
factor as the ratio of the predicted and observed QPO frequencies
(see, for example, Dewangan et al. 2006; Titarchuk \& Fiorito 2004).
Our spectral analysis of X-1 found a $90 \%$ confidence lower limit of
$\Gamma = 2.48$ for the photon index of the power-law component (see
Table 2). While there is a fair amount of scatter in the observed
correlation it is nevertheless evident that no system with a power-law
energy slope this high has a QPO frequency less than 2 Hz (see
Vignarca et al. 2003).  This would suggest a mass scaling of $2 / 0.02
> 100$ for X-1. The lowest accurately measured mass among the relevant
Galactic systems is that of 1655 ($\approx 6.3 M_{\odot}$, Greene,
Bailyn \& Orosz 2001; Shahbaz 2003), which would suggest a lower limit
to the mass of X-1 as $M_{X-1} > 630 M_{\odot}$.  If we instead scale
using the correlation measured for 1550, which also has a well
determined mass ($\approx 10 M_{\odot}$, Orosz et al. 2002), we find
that its QPO frequency is in the range 3 - 7 Hz when the power-law
photon index is within the 1$\sigma$ range we deduced for X-1.  This
gives a mass range from 1500 - 3500 $M_{\odot}$ for X-1.
Interestingly, the correlation between disk temperature and QPO
frequency deduced for XTE J1550-564 (Sobczak et al. 2000) gives a
temperature range from $\approx 0.55 - 0.7$ keV for the same QPO
frequency range (the inferred inner disk radius ranges from $\approx
30 - 125$ km, Sobczak et al. 2000).  Using the mass - disk temperature
scaling expected for standard accretion disks, $kT_{disk} \propto
M^{-1/4}$, we find a mass range from $(0.55 \; {\rm keV} / 0.15 \;
{\rm keV})^4 = 1810$ to $(0.7 \; {\rm keV} / 0.15 \; {\rm keV})^4 =
4740 M_{\odot}$ that overlaps substantially with the scaling deduced
from the QPO frequency and power-law slope, that is, {\it both} the
cool disk temperature and QPO frequency would appear to favor a IMBH
interpretation for X-1.  

What about the luminosity?  When 1550 shows QPOs in the 3 - 7 Hz
frequency range its X-ray luminosity (2 - 20 keV) is about $2 - 3
\times 10^{38} \; (d / 6 \; {\rm kpc})^2$ ergs s$^{-1}$. Taking
typical spectral parameters appropriate to 1550, and scaling into the
0.3 - 10 keV band, we obtain a luminosity over the same energy band
that is a factor of $\approx 15$ less than the observed luminosity of
X-1 (assuming 1550 is 6 kpc distant). This is less than the mass scale
factors inferred above from the QPO frequency and disk temperatures,
however, we do not know {\it a priori} exactly what energy bands in
which to compare luminosities.  Put another way, we don't know
accurately the bolometric corrections.  Nevertheless, a rough
comparison of X-ray luminosities would suggest a mass of $\approx 150
M_{\odot}$ for X-1 if it is radiating at the same Eddington ratio as
1550.  This is about a factor of 10 less than the mass estimates
deduced from the comparison of QPO frequencies above.

While the above arguments appear reasonable to us, it is possible that
the correlation between QPO frequency and power-law index is not
``universal.''  If this is the case, then deriving a mass scaling from
the properties of the Galactic systems becomes more problematic. In
particular, in order to deduce an accurate mass scaling based on
observed QPO frequencies we should require that the accretion disks in
both the Galactic BHs and ULXs extend to approximately the same
relative radius (in terms of $GM/c^2$).  For example, if the disk
radius which dominates the observed ``cool'' disk component in NGC
5408 is relatively farther out than the location of the inner disk
during observations of a Galactic BH with which it is being compared,
then we could be overestimating the mass scale factor.  Soria,
Goncalves \& Kuncic (2006) outline a ``chilled disk'' scenario for
ULXs in which some fraction of their inner disk power is transferred
to a corona and not radiated directly, and suggest that it might
obviate the need for IMBHs in some ULXs. We will explore in detail the
implications of this model for the mass of NGC 5408 X-1 in a
subsequent paper.
 
In an attempt to be conservative we can try to identify a minimum mass
for X-1 based on scaling from the lowest frequency QPOs {\it ever}
observed from a particular Galactic source.  For example, XTE
J1550-564 has shown QPOs with frequencies at $\approx 0.1$ Hz,
although we note that the power-law index was much flatter
(approaching 1.5), and the luminosity lower at such times.  Scaling to
this frequency would indicate a factor of 5 in mass, suggesting a
minimum mass for X-1 near 50 $M_{\odot}$, still more massive than
anything known in the Galaxy, but perhaps consistent with stellar
evolution of low metallicity objects (Fryer \& Kalogera 2001; Heger \&
Woosley 2002; Yungelson 2006).  A source which has shown QPOs with
frequencies near 20 mHz is 1915 (Morgan, Remillard \& Greiner 1999).
If the QPOs in X-1 are analogous to these lower frequency QPOs in
1915, then that would tend to argue against the need for a high mass.
However, the properties of 1915 have been studied by several authors,
and they conclude that these very low frequency QPOs appear to be
associated with the oscillations from and to different emission states
exhibited by the source (see Belloni et al. 2000).  Moreover, 1915
does show higher frequency QPOs in both very high states and low hard
states which fit into the timing and spectral correlations discussed
above. Since X-1 does not show any strong evidence for similar rapid
state changes, it seems more natural to us to associate the QPOs in
X-1 with the $> 1$ Hz QPOs in 1915.

The evidence for a pair of sharp, closely spaced QPOs in X-1 is
somewhat suggestive of the high frequency QPO pairs seen in Galactic
systems (Remillard et al.  2002; Strohmayer 2001; McClintock \&
Remillard 2006). The frequencies of QPO pairs in Galactic systems seem
to prefer a 3:2 ratio (Abramowicz \& Kluzniak 2001; Remillard et
al. 2002).  When the QPO pair in X-1 is most prominent (see Figure 4)
their centroids are accurately measured (see Table 1), and they are
inconsistent with a 3:2 ratio at a little more than the 3$\sigma$
level.  They are also inconsistent with a simple harmonic
relationship, but they are consistent with a 4:3 ratio. As noted
above, one possibility is that we detect two QPO types (for example
type-C and -B) in a manner similar to that sometimes observed in
Galactic systems (Casella et al. 2004). However, with just the one
example so far it is hard to know whether this has any important
physical meaning.  If X-1 is really a $\sim 1000 M_{\odot}$ black
hole, then we would expect the analogs of the high frequency QPOs in
Galactic systems to appear in the vicinity of 1 - 2 Hz, but we do not
detect any significant variability above 0.5 Hz.  The upper limit
(90\% confidence) to any QPO signal power with a coherence $Q \approx
10$ in the 0.5 - 2 Hz band is $\approx 5\%$ (rms).  This is greater
than the typical broad energy band amplitudes found in the Galactic
high frequency QPOs, but it is approaching the interesting sensitivity
range.  Several additional long pointings at X-1 would bring the limit
near the 1\% (rms) level where Galactic QPOs are often seen.  Such a
detection could solidify the IMBH hypothesis for NGC 5408 X-1 (see,
for example, Abramowicz et al. 2004).

In summary, our XMM-Newton observations have revealed that NGC 5408
X-1 exhibits X-ray timing and spectral properties analogous to those
exhibited by Galactic stellar-mass BHs in the ``very high'' or ``steep
power-law'' state, but with the characteristic variability timescales
(QPO and break frequencies) and disk temperature consistently scaled
down. We infer a characteristic size for the X-ray emitting region
$\sim 100$ times larger than typical inner-disk radii of Galactic BHs.
If such a factor is entirely due to a higher BH mass, it implies an
IMBH with $M \sim 1000 M_{\odot}$. However, a straightforward
comparison of X-ray luminosities suggests a mass closer to $100
M_{\odot}$. The luminosity and timing data would be consistent with
each other if the inner disk radius is also an order of magnitude
larger than in Galactic BHs, in dimensionless units, as we shall
discuss in a follow-up paper. While it remains conceivable that the
object is a stellar-mass BH of exceptionally high mass we can think of
no Galactic system that shows {\it all} of the properties it exhibits.
Further observations should be able to determine if it shows X-ray
timing and spectral correlations similar to the Galactic systems, and
should enable tighter estimates on its mass.


We thank the anonymous referee for a careful review, and for pointing
out the possibility that the QPO pair in X-1 may result from the
observation of different QPO types, as sometimes seen in Galactic BHs.

\clearpage


\begin{deluxetable}{ccccc}
\tablecaption{Results of Power Spectral Modeling for NGC 5408 
X-1\tablenotemark{1}}
\tablehead{\colhead{Parameters} & \colhead{All pn (Fig. 2)} & pn, I2
(Fig. 3) & pn, 33 ksec, I2 (Fig. 4) & All pn+MOS $> 2$keV } \startdata
 
A\tablenotemark{a} & $1.57 \pm 0.7$ & $1.2 \pm 0.6$ & $0.6 \pm 0.3$ &
$0.3 \pm 0.1$ \\[6pt]

$\alpha_1$\tablenotemark{b} & $0.046 \pm 0.1$ & $0.08 \pm 0.1$ & $0.2
\pm 0.1$ & $0.13 \pm 0.06$ \\[6pt]

$\alpha_2$\tablenotemark{c} & $1.6 \pm 0.3$ & $1.5 \pm 0.4$ & $1.4 \pm
0.4$ & $1.55 \pm 0.4$ \\

$\nu_{{\rm break}}$ (mHz)\tablenotemark{d} & $3.5 \pm 0.3$ & $4.8 \pm
0.6$ & $4.6 \pm 0.8$ & $25.2 \pm 4.5$ \\

$N_1$\tablenotemark{e} & $0.7 \pm 0.2$ & $0.97 \pm 0.3$ & $1.65 \pm
0.7$ & NA \\

$\nu_1$ (mHz)\tablenotemark{f} & $11.4 \pm 0.7$ & $13.9 \pm 0.5$ &
$14.9 \pm 0.4$ & NA \\

$\sigma_1$ (mHz)\tablenotemark{g} & $5.4\pm 2.5$ & $3.0\pm 1.7$ &
$2.0\pm 1.0$ & NA \\

$N_2$ & $1.23 \pm 0.2$ & $1.81 \pm 0.2$ & $1.99 \pm 0.4$ & 2.1 \\

$\nu_2$ (mHz) & $19.8 \pm 0.2$ & $20.5 \pm 0.3$ & $20.2 \pm 0.3$ & 20.5 \\

$\sigma_2$ (mHz) & $3.4\pm 1.3$ & $2.7\pm 0.6$ & $2.4\pm 0.7$ & 0.4 \\

$N_3$ & $0.5 \pm 0.15$ & $0.57 \pm 0.2$ & $0.82 \pm 0.5$ & NA \\

$\nu_3$ (mHz) & $27.7 \pm 2.0$ & $27.0 \pm 0.8$ & $26.2 \pm 0.9$ & NA \\

$\sigma_3$ (mHz) & $10.4\pm 4.5$ & $3.0\pm 2.0$ & $2.0\pm 1.5$ & NA \\

$\chi^2$ (dof) & 102.2 (118) & 105.1 (118) & 97.0 (118)  & 152.2 (160) \\

\enddata 
\tablenotetext{1}{Summary of best fit power spectral models for NGC 5408 X-1. 
The results from fits to four different power spectra are shown in columns 2-5.
Columns 2-4 show results using pn data over the full energy band, but for
different time intervals. The particular time interval used is given with a 
reference in the heading to the figure the power spectrum appears in.  These
fits used three Lorentzian components, numbered 1-3 in order of increasing
frequency. The last column shows results from a sum of pn and MOS data for 
$> 2$ keV photons, and only one Lorentzian component was used (at 20 mHz).} 
\tablenotetext{a}{Normalization of the broken power law component.}
\tablenotetext{b}{Power law index below the break frequency.}
\tablenotetext{c}{Power law index above the break frequency.}
\tablenotetext{d}{Break frequency, in mHz.}
\tablenotetext{e}{Normalization of the lowest frequency QPO
component.}  \tablenotetext{f}{Centroid frequency, in mHz, of the
lowest frequency QPO component.}  \tablenotetext{g}{Width, in mHz, of
the lowest frequency QPO component.}

\end{deluxetable}

\clearpage
\begin{deluxetable}{cc}
\tablecaption{Spectral Fits to XMM-Newton pn Spectrum\tablenotemark{*}
\label{tbl-2}}
\tablehead{
\colhead{Spectral parameters} &  \colhead{} 
}
\startdata
\multicolumn{2}{c}{Model: {\tt tbabs}*({\tt diskpn} + {\tt pow})} \\
n$_H$\tablenotemark{a} & 6.5$^{+0.04}_{-0.10}$ \\ 
T$_{max}$\tablenotemark{b} & 0.176$^{+0.013}_{-0.004}$ \\
$\Gamma$ & 2.56$^{+0.05}_{-0.08}$ \\
$\chi^2$/dof & 818.1/668 \\
F$_X$ (0.3-10\,keV)\tablenotemark{c} & 3.02$\times10^{-12}$ \\
\hline
\multicolumn{2}{c}{Model: {\tt tbabs}*({\tt diskpn} + {\tt apec} 
+ {\tt pow})} \\
n$_H$\tablenotemark{a} & 7.0$^{+0.08}_{-0.06}$ \\
T$_{max}$\tablenotemark{b} & 0.147$^{+0.014}_{-0.01}$ \\
kT\tablenotemark{d} & 0.873$^{+0.06}_{-0.05}$ \\
$\Gamma$ & 2.56$^{+0.03}_{-0.05}$ \\
$\chi^2$/dof & 682.3/666 \\
F$_X$ (0.3-10\,keV)\tablenotemark{c} & 3.15$\times10^{-12}$ \\
\enddata
\tablenotetext{a}{Hydrogen column density in units of 10$^{20}$\,cm$^{-2}$, 
not including the Galactic contribution of n$_{H Gal} = 
5.73\times10^{20}$\,cm$^{-2}$.}
\tablenotetext{b}{Disk temperature in keV from the XSPEC disk model 
{\tt diskpn}.
The inner disk radius was fixed at 6\,GM/c$^2$.}
\tablenotetext{c}{Unabsorbed flux in units of erg\,cm$^{-2}$\,s$^{-1}$.}
\tablenotetext{d}{Plasma temperature in keV from the XSPEC model {\tt apec}.  
The abundances were fixed to the solar values.}
\tablenotetext{*}{All errors are quoted at the 90\% confidence level.}
\end{deluxetable}

\clearpage

\begin{figure}
\begin{center}
 \includegraphics[width=6.5in, height=6in]{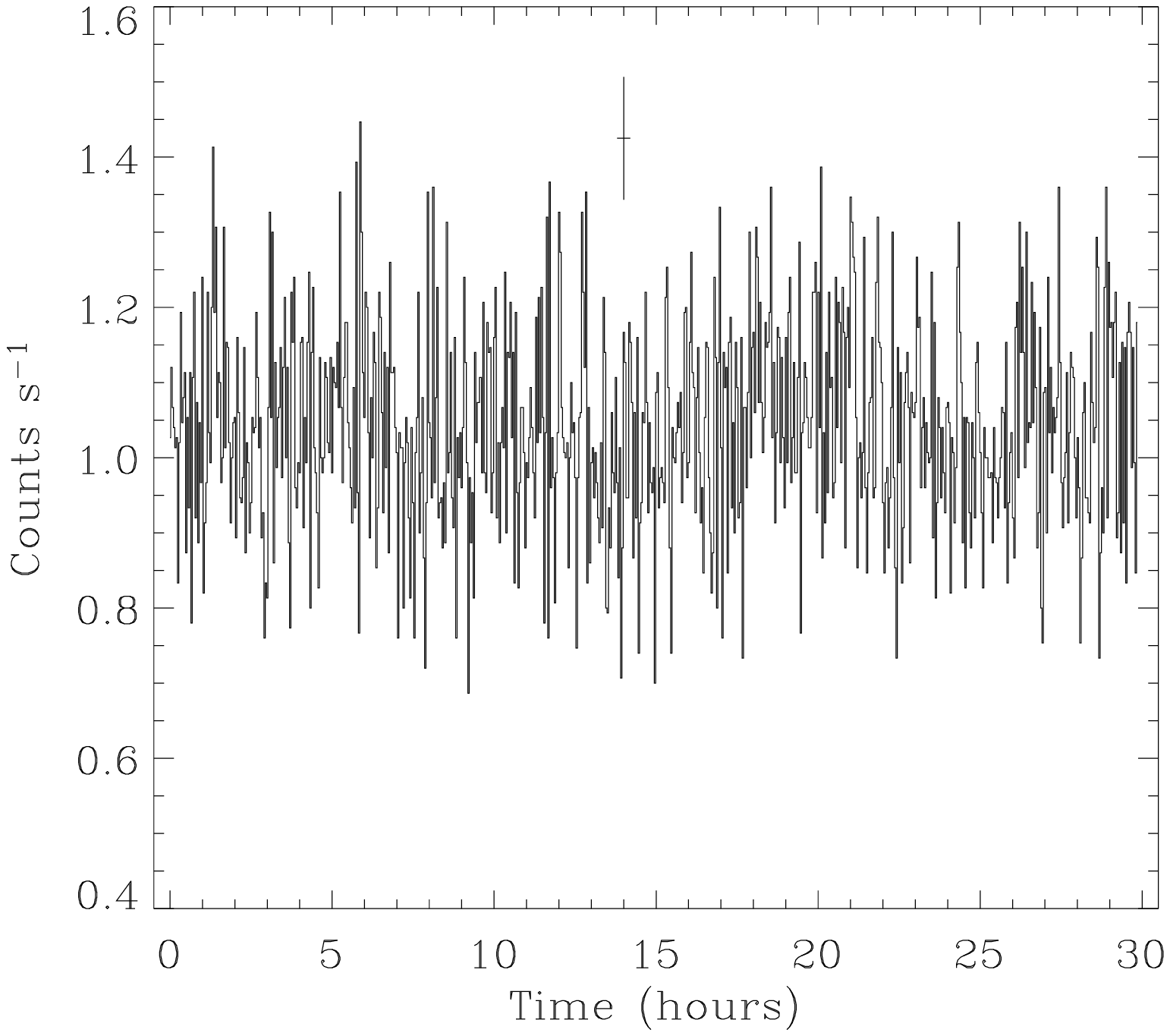}
\end{center}
Figure 1: Lightcurve of NGC 5408 X-1 (0.2 - 15 keV band) from
XMM-Newton EPIC observations.  The count rate shown is the sum of the
pn and MOS cameras.  The bin size is 150 seconds. Time zero
corresponds to MJD 53748.8378092 (TT).  A characteristic error bar is
also shown.
\end{figure}

\clearpage

\begin{figure}
\begin{center}
 \includegraphics[width=6.5in, height=6in]{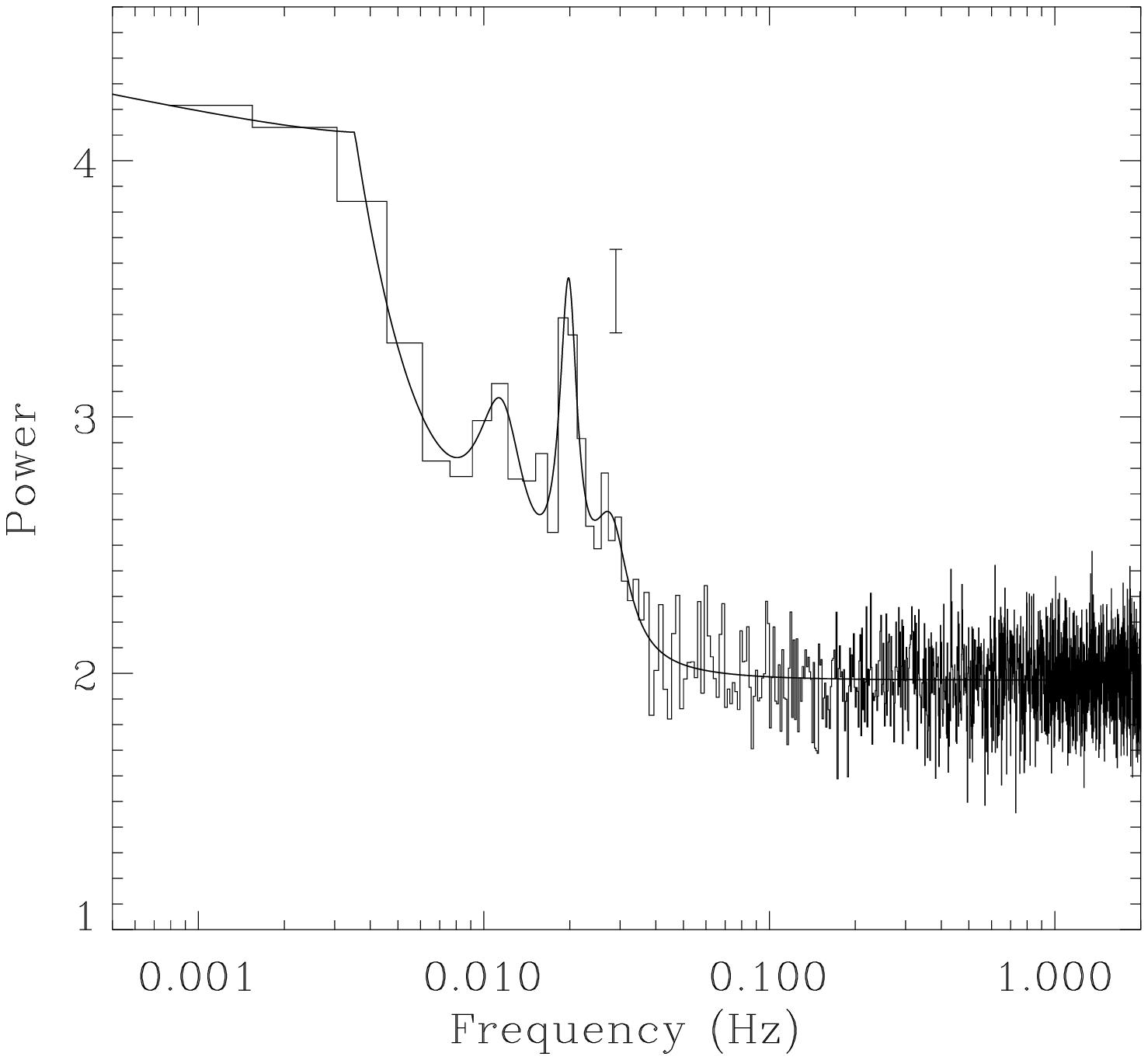}
\end{center}
Figure 2: Average power spectrum of NGC 5408 X-1 from all the EPIC/pn
data (histogram) and the best fitting model (solid). The frequency
resolution is 1.52 mHz, and each bin is an average of 150 independent
power spectral measurements.  The effective exposure is $\approx 100$
ksec. A characteristic error bar is also shown. See the text for a
detailed discussion of the model, and Table 1 for model parameters. 
\end{figure}

\clearpage

\begin{figure}
\begin{center}
 \includegraphics[width=6.5in, height=6in]{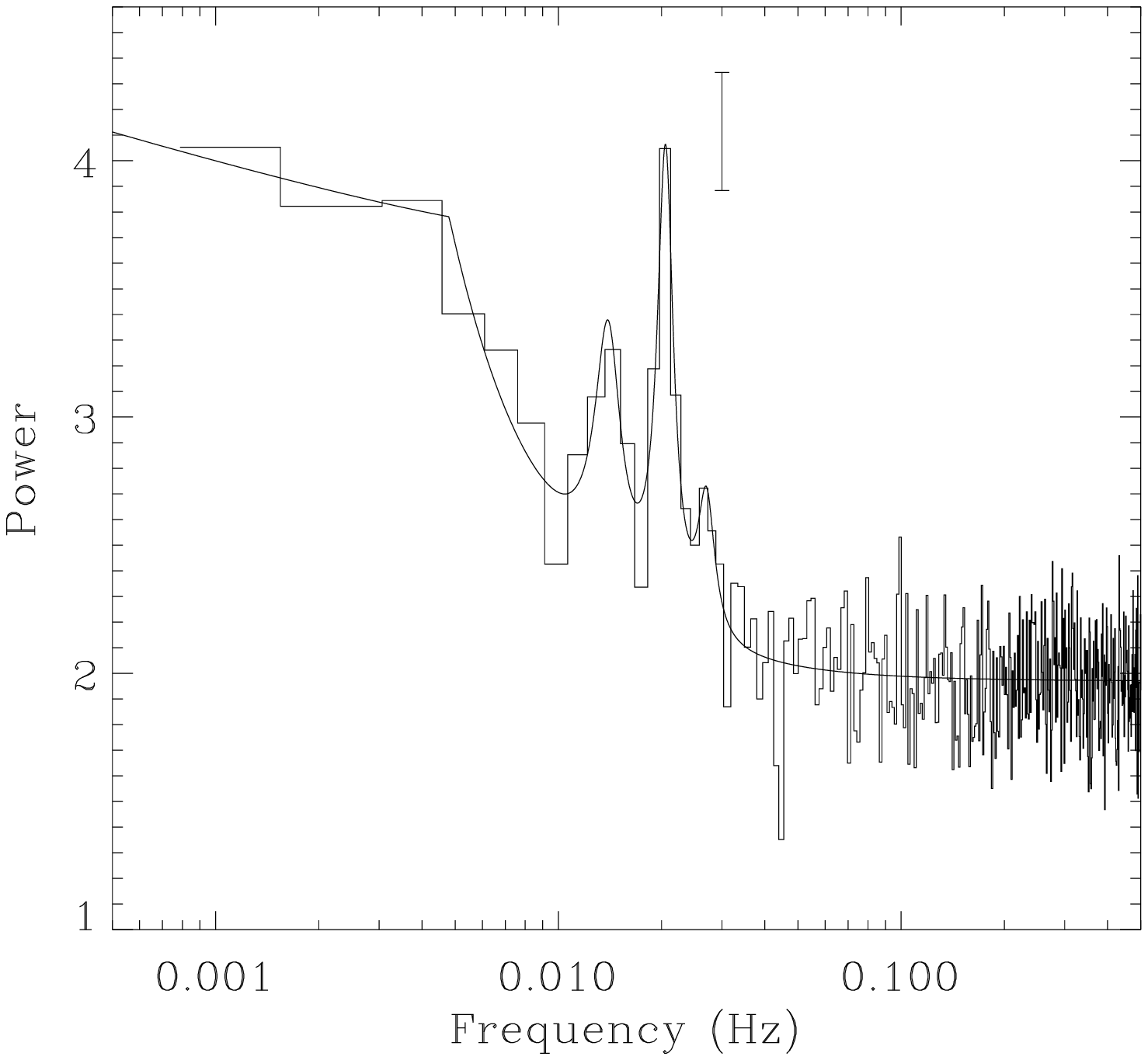}
\end{center}
Figure 3: Average power spectrum of NGC 5408 X-1 from EPIC/pn data
(histogram) and the best fitting model (solid). Here we used data only
from the longest continuous good time interval ($\approx 57$
ksec). The frequency resolution is 1.52 mHz, and each bin is an
average of 75 independent power spectral measurements. A
characteristic error bar is also shown. See the text for a detailed
discussion of the model, and Table 1 for model parameters.

\end{figure}

\clearpage

\begin{figure}
\begin{center}
 \includegraphics[width=6.5in, height=6in]{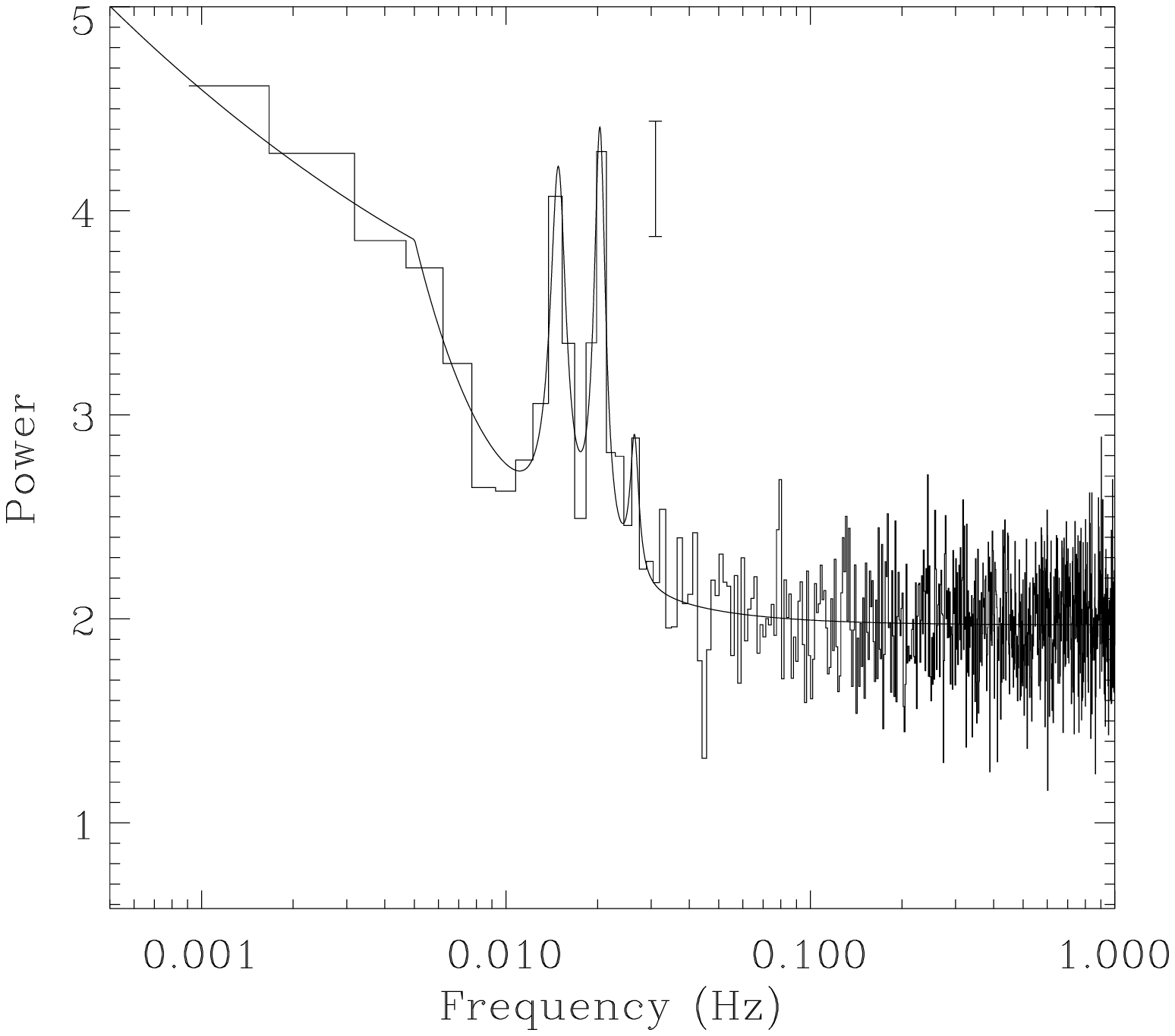}
\end{center}
Figure 4: Average power spectrum of NGC 5408 X-1 from EPIC/pn data
(histogram) and the best fitting model (solid). This spectrum was
constructed from the first 33 ksec of the longest continuous time
interval (interval 2). The frequency resolution is 1.52 mHz, and each
bin is an average of 50 independent power spectral measurements. A
characteristic error bar is also shown. See the text for a detailed
discussion of the model, and Table 1 for model parameters.
\end{figure}
\clearpage

\begin{figure}
\begin{center}
 \includegraphics[width=6.5in, height=6in]{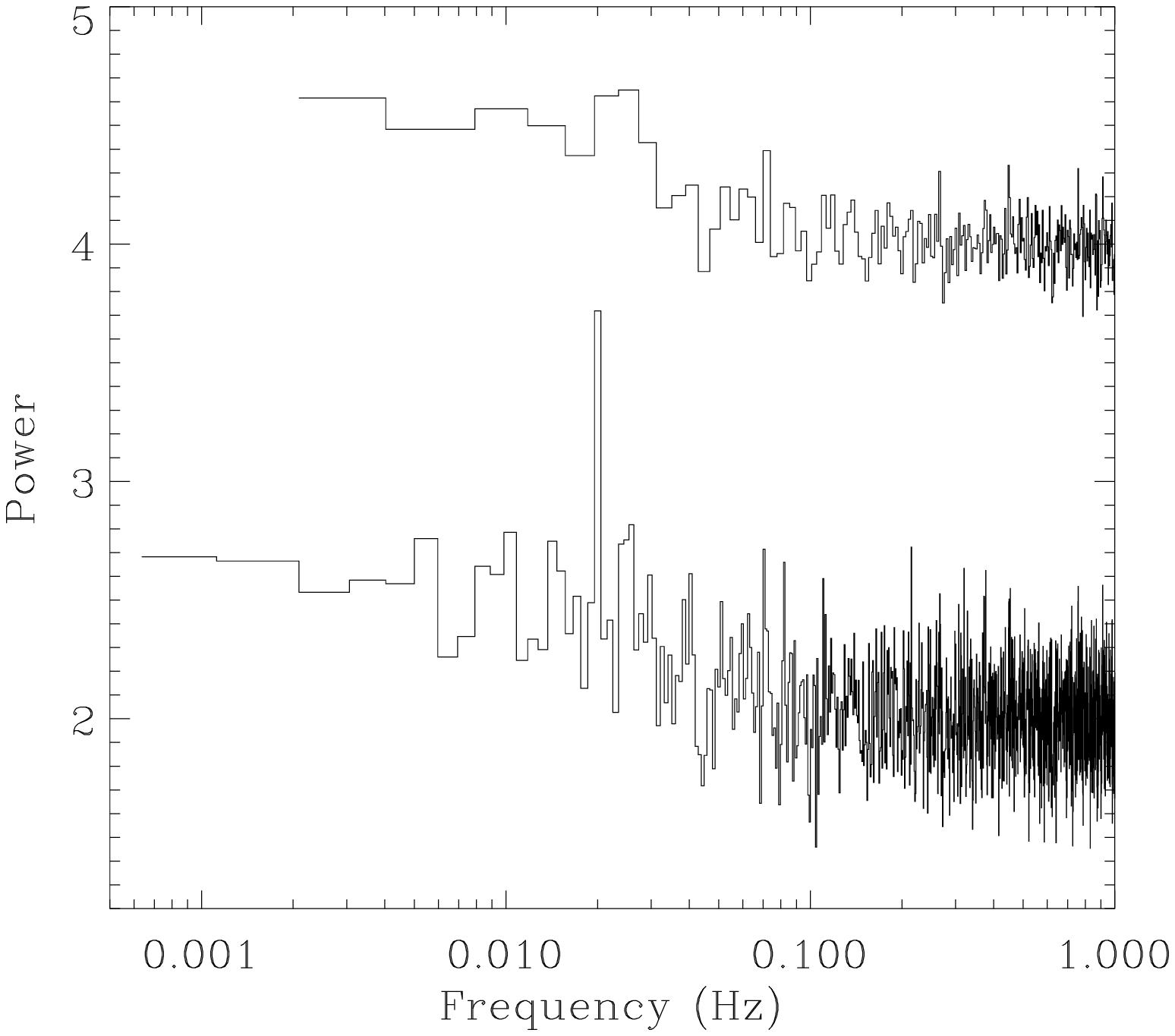}
\end{center}
Figure 5: Average power spectrum of NGC 5408 X-1 from all EPIC/pn+MOS
data. Here we used only photons with energy $> 2$ keV. The bottom and
top curves show the same data but rebinned to different frequency
resolutions of 0.97 and 3.88 mHz, respectively.
\end{figure}
\clearpage

\begin{figure}
\begin{center}
 \includegraphics[width=6.5in, height=6in]{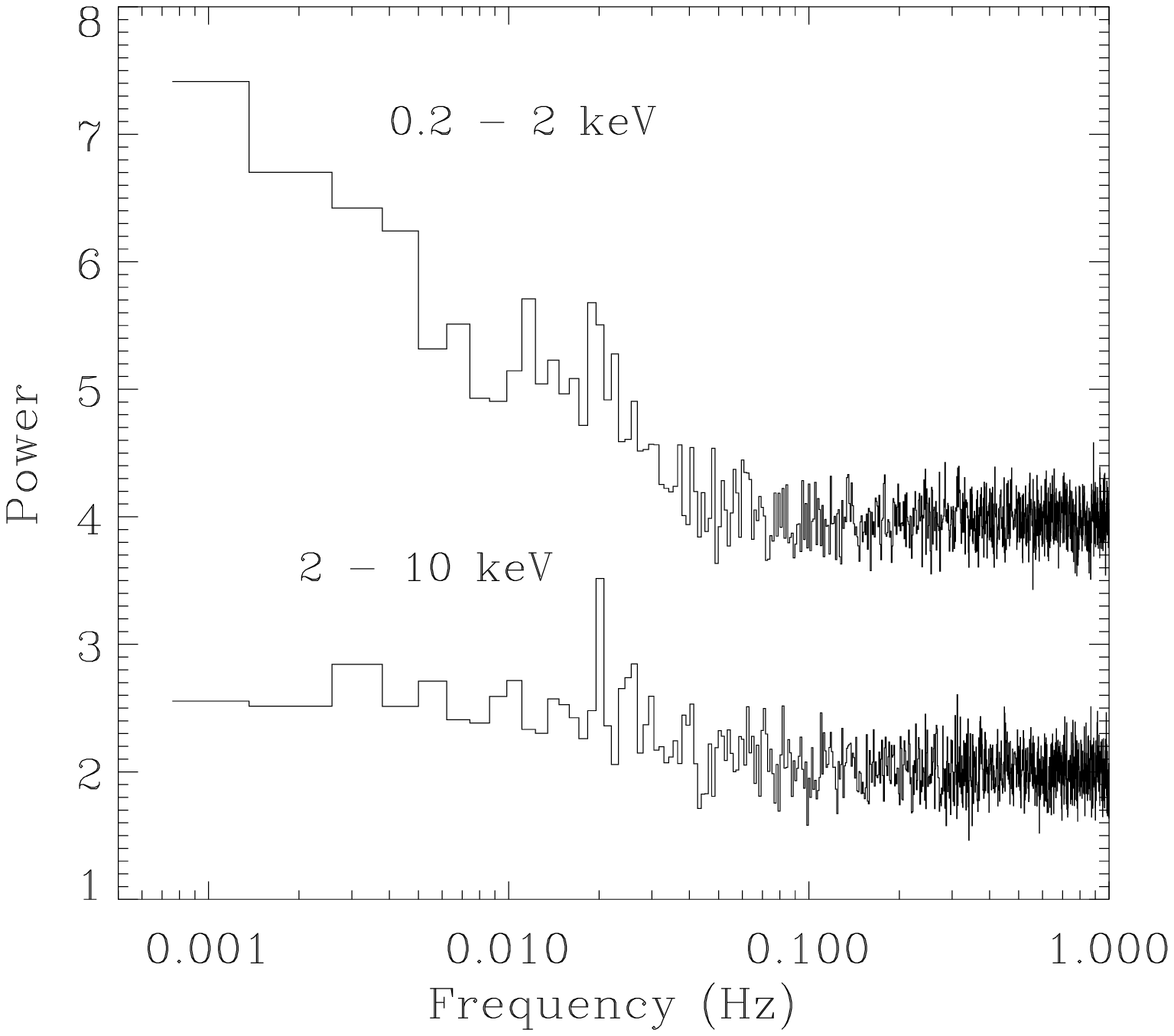}
\end{center}
Figure 6: Comparison of average power spectra of NGC 5408 X-1 in two
energy bands from EPIC/pn+MOS data. The frequency resolution is 1.2
mHz in both spectra. The upper curve is displaced vertically by 2 for
clarity.
\end{figure}
\clearpage

\begin{figure}
\begin{center}
 \includegraphics[width=6.5in, height=6in]{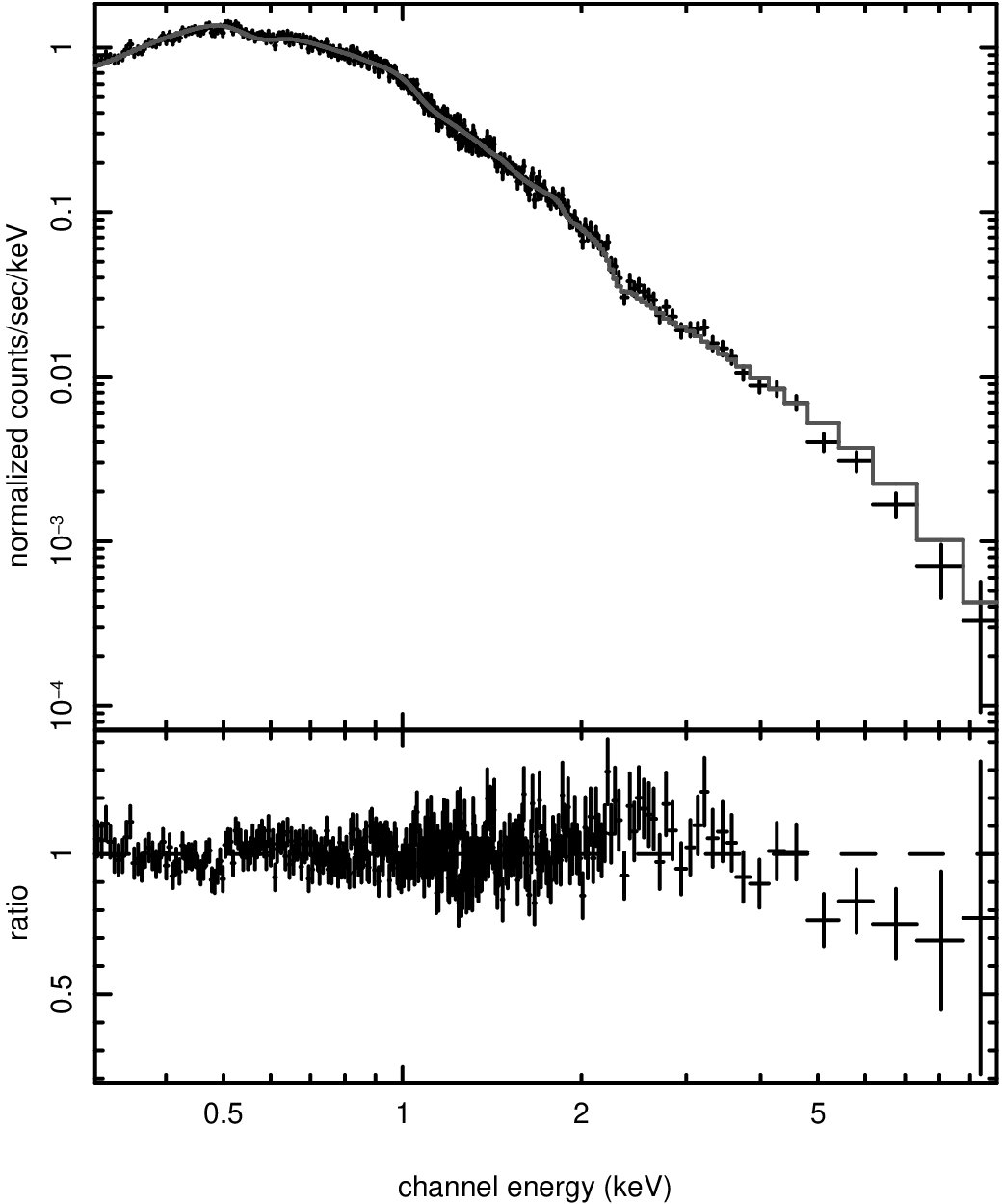}
\end{center}
Figure 7: Energy spectrum of NGC 5408 X-1 from EPIC/pn data.  The top
panel shows the count rate spectrum (data points with error bars) and
the best fitting model (solid histogram). The bottom panel shows the
ratio of data to model.  See Table 2 for details of the spectral
model.
\end{figure}


\begin{thebibliography}

\bibitem[Abramowicz et al.(2004)]{2004ApJ...609L..63A} Abramowicz,
M.~A., Klu{\'z}niak, W., McClintock, J.~E., \& Remillard, R.~A.\ 2004,
ApJ, 609, L63.

\bibitem[Abramowicz \& Klu{\'z}niak(2001)]{2001A&A...374L..19A}
Abramowicz, M.~A., \& Klu{\'z}niak, W.\ 2001, A\&A, 374, L19.

\bibitem[Begelman(2006)]{2006ApJ...643.1065B} Begelman, M.~C.\ 2006,
ApJ, 643, 1065.

\bibitem[Belloni et al.(2000)]{2000A&A...355..271B} Belloni, T.,
Klein-Wolt, M., M{\'e}ndez, M., van der Klis, M., \& van Paradijs, J.\
2000, A\&A, 355, 271.

\bibitem[Belloni et al.(1996)]{1996ApJ...472L.107B} Belloni, T.,
Mendez, M., van der Klis, M., Hasinger, G., Lewin, W.~H.~G., \& van
Paradijs, J.\ 1996, ApJ, 472, L107.

\bibitem[Belloni et al.(1997)]{1997A&A...322..857B} Belloni, T., van
der Klis, M., Lewin, W.~H.~G., van Paradijs, J., Dotani, T., Mitsuda,
K., \& Miyamoto, S.\ 1997, \aap, 322, 857.

\bibitem[Casella et al.(2004)]{2004A&A...426..587C} Casella, P.,
Belloni, T., Homan, J., \& Stella, L.\ 2004, \aap, 426, 587.

\bibitem[Casella et al.(2005)]{2005ApJ...629..403C} Casella, P.,
Belloni, T., \& Stella, L.\ 2005, \apj, 629, 403.

\bibitem[Colbert \& Mushotzky (1999)]{CM99}Colbert, E. J. M. \&
Mushotzky, R.  F. 1999, ApJ, 519, 89.

\bibitem[Cropper et al.(2004)]{2004MNRAS.349...39C} Cropper, M.,
Soria, R., Mushotzky, R.~F., Wu, K., Markwardt, C.~B., \& Pakull, M.\
2004, MNRAS, 349, 39.

\bibitem[Dewangan et al.(2006)]{2006ApJ...637L..21D} Dewangan, G.~C.,
Titarchuk, L., \& Griffiths, R.~E.\ 2006, ApJ, 637, L21.

\bibitem[Dewangan et al.(2004)]{2004ApJ...608L..57D} Dewangan, G.~C.,
Miyaji, T., Griffiths, R.~E., \& Lehmann, I.\ 2004, ApJ, 608, L57.

\bibitem[Fabbiano \& White (2006)]{FW06}Fabbiano, G. \& White,
N. E. 2006, in ``Compact Stellar X-ray Sources,'' ed. W. H. G. Lewin,
\& M. van der Klis, (Cambridge University Press: Cambridge), pg. 475.

\bibitem[Feng \& Kaaret(2005)]{2005ApJ...633.1052F} Feng, H., \& Kaaret, 
P.\ 2005, \apj, 633, 1052.

\bibitem[Fiorito \& Titarchuk(2004)]{2004ApJ...614L.113F} Fiorito, R.,
\& Titarchuk, L.\ 2004, ApJ, 614, L113.

\bibitem[Frank, King \& Raine (2002)]{FKR02}Frank, J., King, A. R. \&
Raine, D.  2002, in ``Accretion Power in Astrophysics,'' (Cambridge
University Press: Cambridge).

\bibitem[Fryer \& Kalogera(2001)]{2001ApJ...554..548F} Fryer, C.~L., \& 
Kalogera, V.\ 2001, ApJ, 554, 548.

\bibitem[Greene et al.(2001)]{2001ApJ...554.1290G} Greene, J., Bailyn,
C.~D., \& Orosz, J.~A.\ 2001, ApJ, 554, 1290.

\bibitem[Heger \& Woosley(2002)]{2002ApJ...567..532H} Heger, A., \& 
Woosley, S.~E.\ 2002, ApJ, 567, 532.

\bibitem[Homan et al.(2001)]{2001ApJS..132..377H} Homan, J., Wijnands, R., 
van der Klis, M., Belloni, T., van Paradijs, J., Klein-Wolt, M., Fender, 
R., \& M{\'e}ndez, M.\ 2001, \apjs, 132, 377.

\bibitem[Kaaret et al.(2006)]{2006ApJ...646..174K} Kaaret, P., Simet, 
M.~G., \& Lang, C.~C.\ 2006, ApJ, 646, 174 

\bibitem[Kaaret et al.(2003)]{2003Sci...299..365K} Kaaret, P., Corbel,
S., Prestwich, A.~H., \& Zezas, A.\ 2003, Science, 299, 365.

\bibitem[Kaaret et al. (2001)]{K01}Kaaret, P. et al. 2001, MNRAS, 321,
L29.

\bibitem[Karachentsev et al.(2002)]{2002A&A...385...21K} Karachentsev,
I.~D., et al.\ 2002, A\&A, 385, 21.

\bibitem[King et al. (2001)]{K2001} King, A. R., Davies, M. B., Ward,
M. J., Fabbiano, G. \& Elvis, M. 2001, ApJ, 552, L109.

\bibitem[Kong et al.(2004)]{2004ApJ...617L..49K} Kong, A.~K.~H., Di 
Stefano, R., \& Yuan, F.\ 2004, ApJ, 617, L49.

\bibitem[Leahy et al.(1983)]{1983ApJ...266..160L} Leahy, D.~A., Darbro, W., 
Elsner, R.~F., Weisskopf, M.~C., Kahn, S., Sutherland, P.~G., \& Grindlay, 
J.~E.\ 1983, ApJ, 266, 160.

\bibitem[Markowitz et al.(2003)]{2003ApJ...593...96M} Markowitz, A.,
et al.\ 2003, ApJ, 593, 96.

\bibitem[McClintock \& Remillard (2006)]{MR06}McClintock, J. E. \&
Remillard, R. A. 2006, in ``Compact Stellar X-ray Sources,''
ed. W. H. G. Lewin, \& M. van der Klis, (Cambridge University Press:
Cambridge), pg. 157.

\bibitem[McHardy et al.(2006)]{2006Natur.444..730M} McHardy, I.~M.,
Koerding, E., Knigge, C., Uttley, P., \& Fender, R.~P.\ 2006, \nat,
444, 730.

\bibitem[Miller \& Colbert (2004)]{MC04}Miller, M.~C., \& Colbert,
E.~J.~M.\ 2004, International Journal of Modern Physics D, 13, 1.

\bibitem[Miller et al.(2004)]{2004ApJ...614L.117M} Miller, J.~M.,
Fabian, A.~C., \& Miller, M.~C.\ 2004a, ApJ, 614, L117.

\bibitem[Miller et al.(2004)]{2004ApJ...607..931M} Miller, J.~M.,
Fabian, A.~C., \& Miller, M.~C.\ 2004b, \apj, 607, 931.

\bibitem[Miller et al. (2003)]{M2003} Miller, J. M., Fabbiano, G.,
Miller, M. C., \& Fabian, A. C. 2003, ApJ, 585, L37.

\bibitem[Morgan, E. H. et al. (1997)]{M97} Morgan, E. H., Remillard,
R. A.  \& Greiner, J. 1997, ApJ, 482, 993.

\bibitem[Mucciarelli et al.(2006)]{2006MNRAS.365.1123M} Mucciarelli,
P., Casella, P., Belloni, T., Zampieri, L., \& Ranalli, P.\ 2006,
MNRAS, 365, 1123.

\bibitem[Muno et al. (1999)]{MMR99} Muno, M. P., Morgan, E. H. \&
Remillard, R. A. 1999, ApJ, 527, 321.

\bibitem[Orosz et al.(2002)]{2002ApJ...568..845O} Orosz, J.~A., et
al.\ 2002, ApJ, 568, 845.

\bibitem[Reig et al.(2003)]{2003A&A...412..229R} Reig, P., Belloni,
T., \& van der Klis, M.\ 2003, A\&A, 412, 229.

\bibitem[Remillard et al.(2002)]{2002ApJ...564..962R} Remillard, R.~A., 
Sobczak, G.~J., Muno, M.~P., \& McClintock, J.~E.\ 2002, \apj, 564, 962.

\bibitem[Remillard et al. (2002)]{Rem02} Remillard, R, A., Muno,
M. P., McClintock, J. E. \& Orosz, J. A. 2002, ApJ, 580, 1030.

\bibitem[Remillard et al. (1999)]{R99} Remillard, R. A., Morgan,
E. H., McClintock, J. E., Bailyn, C. D. \& Orosz, J. A. 1999, ApJ,
522, 397.

\bibitem[Shahbaz(2003)]{2003MNRAS.339.1031S} Shahbaz, T.\ 2003, MNRAS,
339, 1031.

\bibitem[Sobczak et al.(2000)]{2000ApJ...531..537S} Sobczak, G.~J.,
McClintock, J.~E., Remillard, R.~A., Cui, W., Levine, A.~M., Morgan,
E.~H., Orosz, J.~A., \& Bailyn, C.~D.\ 2000, \apj, 531, 537.

\bibitem[Soria et al.(2006a)]{2006IAUS..238E.162S} Soria, R., Goncalves, 
A.~C., \& Kuncic, Z.\ 2006, IAU Symposium, 238, 162.

\bibitem[Soria et al.(2006)]{2006MNRAS.370.1666S} Soria, R., Kuncic, Z., 
Broderick, J.~W., \& Ryder, S.~D.\ 2006, MNRAS, 370, 1666.

\bibitem[Soria et al.(2006)]{2006MNRAS.368.1527S} Soria, R., Fender,
R.~P., Hannikainen, D.~C., Read, A.~M., \& Stevens, I.~R.\ 2006,
MNRAS, 368, 1527.

\bibitem[Soria et al.(2004)]{2004A&A...423..955S} Soria, R., Motch,
C., Read, A.~M., \& Stevens, I.~R.\ 2004, A\&A, 423, 955.

\bibitem[Stobbart et al.(2006)]{2006MNRAS.368..397S} Stobbart, A.-M., 
Roberts, T.~P., \& Wilms, J.\ 2006, MNRAS, 368, 397.

\bibitem[Strohmayer \& Mushotzky(2003)]{2003ApJ...586L..61S}
Strohmayer, T.~E., \& Mushotzky, R.~F.\ 2003, ApJ, 586, L61.

\bibitem[Strohmayer(2001)]{2001ApJ...552L..49S} Strohmayer, T.~E.\
2001, ApJ, 552, L49.

\bibitem[Titarchuk \& Fiorito(2004)]{2004ApJ...612..988T} Titarchuk,
L., \& Fiorito, R.\ 2004, ApJ, 612, 988.

\bibitem[Vignarca et al.(2003)]{2003A&A...397..729V} Vignarca, F.,
Migliari, S., Belloni, T., Psaltis, D., \& van der Klis, M.\ 2003,
A\&A, 397, 729.

\bibitem[Wijnands et al.(1999)]{1999ApJ...526L..33W} Wijnands, R., Homan, 
J., \& van der Klis, M.\ 1999, \apjl, 526, L33.

\bibitem[Winter et al.(2006)]{2006ApJ...649..730W} Winter, L.~M., 
Mushotzky, R.~F., \& Reynolds, C.~S.\ 2006, ApJ, 649, 730


\end{thebibliography}
\end{document}